\documentclass[aps,prl,showpacs]{revtex4}
\usepackage{graphicx}

\begin{document}

\title{Quantum parallelism of the controlled-NOT operation:\\
an experimental criterion for the evaluation of device performance}


\author{Holger F. Hofmann}
\email{h.hofmann@osa.org}
\affiliation{
Graduate School of Advanced Sciences of Matter, Hiroshima University,
Kagamiyama 1-3-1, Higashi Hiroshima 739-8530, Japan}

\begin{abstract}

It is shown that a quantum controlled-NOT gate simultaneously
performs the logical functions of three distinct conditional
local operations. Each of these local operations can be verified
by measuring a corresponding truth table of four local inputs
and four local outputs. The quantum parallelism of the gate can
then be observed directly in a set of three simple experimental
tests, each of which has a clear intuitive interpretation in
terms of classical logical operations. Specifically, quantum
parallelism is achieved if the average fidelity of the three
classical operations exceeds 2/3. It is thus possible to evaluate
the essential quantum parallelism of an experimental
controlled-NOT gate by testing only three characteristic classical
operations performed by the gate.
\end{abstract}

\pacs{
03.67.Lx 
03.67.Mn 
03.65.Yz 
03.65.Ud 
}

\maketitle

Quantum information science may provide a wide range
of new technologies by making the unique properties of the
quantum world available for the processing and transmission
of data. In particular, quantum computation may
enhance the efficiency of computational tasks by exploiting the
quantum parallelism of quantum logic operations.
In order to realize such an efficient quantum computer, it is
necessary to implement a universal set of quantum gates,
including at least one interaction between pairs of qubits.
Since the quantum controlled-NOT can provide this essential
interaction, the experimental realization of the controlled-NOT
operation on pairs of qubits is a significant step towards the
realization of universal quantum computation \cite{Intro}.

Recently, there have been several successful demonstrations of
experimental quantum
controlled-NOT gates using superconducting charge qubits \cite{Pas03},
trapped ions \cite{SKa03,Lei03}, and photonic qubits \cite{Bri03,Hua04,Zha05}. However,
each of these realizations has its own characteristic noise signatures,
and the actual performance of the gates is different from the ideal
case of a fully coherent quantum controlled-NOT operation. A complete
characterization of the noise signature of a two qubit gate can only
be achieved by quantum process tomography, which characterizes the
two qubit operation in terms of 256 combinations of input and output
states \cite{Poy97,Chu97,Nie00,Bri04}. Obviously, the experimental effort
involved in such a characterization is very great. It may therefore be
useful to identify the essential operations of the quantum controlled-NOT
in order to define more efficient tests for experimentally realized
quantum gates.

In this paper, it is shown that the ideal quantum controlled-NOT operation can be expanded in terms of a set of three local
operations, minus a dephasing term. Each of the local operations
can be tested using a single
setting of four orthogonal input states and four orthogonal measurement
projections in the output. It is thus possible to characterize the
essential elements of the quantum controlled-NOT gate by measuring the
fidelity of only three classical truth tables. The quantum properties
of the gate can then be identified with the parallel performance of
three well-defined classical logical operations observable in three
different basis sets of distinguishable input and output states.

The unitary operator describing an ideal quantum controlled-NOT operation
can be expressed in a basis-independent manner by using the Pauli
matrices $\{ I, X, Y, Z\}$, where the logical states of the
computational basis are defined by $Z \mid \! 0 \rangle =
\mid \! 0 \rangle$
and $Z \mid \! 1 \rangle = - \mid \! 1 \rangle$ \cite{Intro}.
The effects of the quantum process on an arbitrary two qubit input
density matrix $\hat{\rho}$ can then be written as
\begin{eqnarray}
\label{eq:qcNOT}
E_{\mbox{\tiny CNOT}} (\hat{\rho}) &=&
\hat{U}_{\mbox{\tiny CNOT}} \; \hat{\rho} \;
\hat{U}_{\mbox{\tiny CNOT}}^\dagger
\nonumber \\[0.2cm]
\mbox{with} &&
\hat{U}_{\mbox{\tiny CNOT}} = \frac{1}{2} \left(I \otimes I
+ I \otimes X + Z \otimes I - Z \otimes X \right).
\end{eqnarray}
Here, the unitary operation $\hat{U}_{\mbox{\tiny CNOT}}$
has been expanded in terms of the shortest
possible sum of local operator products \cite{schmidt}.
In this representation, the elementary operations appear to be the
spin flips represented by $X$,$Y$, and $Z$. However, an incoherent
mixture of the four components in equation (\ref{eq:qcNOT}) would
simply result in dephasing between the $Z$ eigenstates in system one
and between the $X$ eigenstates in system two,
\begin{eqnarray}
\label{eq:dephase}
D(\hat{\rho}) &=& \frac{1}{4} \big(
(I \otimes I) \; \hat{\rho} \; (I \otimes I)
+ (I \otimes X) \; \hat{\rho} \; (I \otimes X)
\nonumber \\ &&
+ (Z \otimes I) \; \hat{\rho} \; (Z \otimes I)
+ (Z \otimes X) \; \hat{\rho} \; (Z \otimes X) \big) .
\end{eqnarray}
The comparison between equations (\ref{eq:qcNOT}) and (\ref{eq:dephase})
indicates that all of the essential features of the quantum
controlled-NOT can be given in terms of the coherences between the
elementary spin flip operations. Therefore, an experimental verification
of the quantum gate operation should focus on the observable effects of
these coherences on local input and output states. In the following,
these effects will be identified by considering local operations that
have the same coherences as the quantum controlled-NOT gate.

In local operations, it is always possible to identify a simple
physical interpretation of the coherence between the spin flip
operations. For example, the superpositions $(I+Z)/2$ and $(I-Z)/2$
represent measurement projections on the eigenstates of $Z$. The
coherence between $I \otimes I$ and $Z \otimes I$ and the
coherence between $I \otimes X$ and $Z \otimes X$ can
therefore be understood in terms of a conditional local
operation $L_1(\hat{\rho})$ given by a measurement of $Z$ in system
one followed by a conditional spin flip operation $X$ in system two
if the result of the measurement in system one is $Z=-1$,
\[
L_1(\hat{\rho}) = \hat{\Pi}_{Z0} \; \hat{\rho} \; \hat{\Pi}_{Z0}^\dagger
+ \hat{\Pi}_{Z1} \; \hat{\rho} \; \hat{\Pi}_{Z1}^\dagger
\]
\\[-1cm]
\[
\hat{\Pi}_{Z0}
= \frac{1}{2}\left( I \otimes I + Z \otimes I \right)
= \mid \!Z\!=\!+\!1 \rangle\langle Z\!=\!+\!1\! \mid \otimes I
\]
\\[-1cm]
\begin{equation}
\hat{\Pi}_{Z1}
= \frac{1}{2}\left( I \otimes X - Z \otimes X \right)
= \mid \!Z\!=\!-\!1 \rangle\langle Z\!=\!-\!1\! \mid \otimes X.
\end{equation}
This operation is in fact already a complete controlled-NOT
operation in the computational basis, performed entirely by
conditional local operations (and thus equivalent to an
interaction by local operations and classical communication).
The coherences between $I \otimes I$ and $Z \otimes I$ and
between $I \otimes X$ and $Z \otimes X$ are therefore sufficient
to define the operation of the quantum controlled-NOT gate
in the computational basis, while the other four coherences
have no effect on the gate performance observed in this basis.

A very similar interpretation can be found for the coherences
between $I \otimes I$ and $I \otimes X$ and between $Z \otimes I$
and $Z \otimes X$. In this case, the roles of system one and
two and the roles of $X$ and $Z$ have simply been exchanged.
The conditional local operation $L_2(\hat{\rho})$ therefore
describes a measurement of $X$ in system two,
followed by a conditional spin flip operation $Z$ in system one
if the result of the measurement in system two is $X=-1$,
\[
L_2(\hat{\rho}) = \hat{\Pi}_{X0} \; \hat{\rho} \; \hat{\Pi}_{X0}^\dagger
+ \hat{\Pi}_{X1} \; \hat{\rho} \; \hat{\Pi}_{X1}^\dagger
\]\\[-1cm]
\[
\hat{\Pi}_{X0}
= \frac{1}{2}\left( I \otimes I + I \otimes X \right)
= I \otimes \mid \!X\!=\!+\!1 \rangle \langle X\!=\!+\!1\! \mid
\]\\[-1cm]
\begin{equation}
\hat{\Pi}_{X1}
= \frac{1}{2}\left( I \otimes X - Z \otimes X \right)
= Z \otimes \mid \!X\!=\!-\!1 \rangle \langle X\!=\!-\!1\! \mid.
\end{equation}
As the symmetry between $L_1$ and $L_2$ suggests, this is also
a complete controlled-NOT operation, performed in the $X$-basis
with reversed roles for the target and the control \cite{Mermin}.
The coherences between $I \otimes I$ and $I \otimes X$ and
between $Z \otimes I$ and $Z \otimes X$ are therefore solely responsible
for the performance of the quantum controlled-NOT gate in the $X$-basis.

Finally, a different interpretation is necessary to identify the
effects of the coherences between $I \otimes I$ and $Z \otimes X$
and between $I \otimes X$ and $Z \otimes I$, since this coherence is
symmetric in the two qubits. Such coherences can be obtained
by performing correlated $\pi/2$-rotations of the spins in the two
systems,
\[
L_3(\hat{\rho}) =\frac{1}{2}\left( \hat{U}_{+\pi/2} \; \hat{\rho} \; \hat{U}_{+\pi/2}^\dagger
+ \hat{U}_{-\pi/2} \; \hat{\rho} \; \hat{U}_{-\pi/2}^\dagger \right)
\]\\[-1cm]
\[
\hat{U}_{+\pi/2}
= \frac{1}{2}\left(I + i Z\right) \otimes \left(I + i X \right)
= \exp[+i \frac{\pi}{4} Z] \otimes \exp[+i \frac{\pi}{4} X]
\]\\[-1cm]
\begin{equation}
\hat{U}_{-\pi/2}
= \frac{1}{2}\left(I - i Z\right) \otimes \left(I - i X \right)
= \exp[-i \frac{\pi}{4} Z] \otimes \exp[-i \frac{\pi}{4} X].
\end{equation}
It might be worth noting that this operation corresponds to the
best possible local approximation of a quantum phase gate operation
in the $Z \otimes X$ basis,
since it results in a total phase change of $\pi$ for the eigenstate
with $Z=+1$ in system one and $X=+1$ in system two, while preserving
the phase of the eigenstate with $Z=-1$ and $X=-1$. The ideal
non-local operation given by $\hat{U}_{\mbox{\tiny CNOT}}$ also preserves
the phases of the other two eigenstates, but the price to be paid
for performing the phase shift by local operations only is the
complete randomization of the phases for $Z=-1$ and $X=+1$, and
for $Z=+1$ and $X=-1$.
However, for the purpose of verifying the operation experimentally,
it is more useful to consider the effect of $L_3$ on eigenstates of
$X \otimes Z$ in the input.
Specifically, the $\pi/2$-rotation around the $Z$-axis
can be verified by using an $X$ eigenstate as input which should be
transformed into the corresponding $Y$ eigenstate in the output.
Likewise, the $\pi/2$-rotation around the $X$-axis can be verified by
using a $Z$ eigenstate as input. Since the operation $L_3$ is a
mixture of two possible rotation directions,
the output states for
this input basis are mixtures of $Y \otimes Y$-eigenstates, with $\langle Y \otimes Y \rangle(\mbox{out}) = - \langle X \otimes Z \rangle (\mbox{in})$.
It is therefore possible to observe the coherences between
$I \otimes I$ and $Z \otimes X$ and between $I \otimes X$ and
$Z \otimes I$ by using eigenstates of $X \otimes Z$ as input states
and by measuring $Y \otimes Y$ in the output.
Note that the reverse is also possible, but this choice of input and
output basis makes it easier to estimate the entanglement capability
of the gate from the fidelities of the observed operations, as will
be explained below.

It is now possible to express the ideal quantum controlled-NOT operation
in terms of the three local operations $L_i$ and the dephasing operation
$D$. This expansion of the quantum process reads
\begin{equation}
\label{eq:expand}
 E_{\mbox{\tiny CNOT}} (\hat{\rho}) =
L_1(\hat{\rho}) + L_2(\hat{\rho}) + L_3(\hat{\rho}) - 2 D(\hat{\rho}).
\end{equation}
The role of the negative dephasing term in this expansion can be
understood by considering the observable effects of the local
operations $L_i$ in different basis settings. As discussed above,
each local operation is
associated with a characteristic logical operation observed
using a specific selection of input and output states.
For each of these specific input and output settings,
the effects of the other two local operations are indistinguishable
from the effects of dephasing. The negative dephasing
term in equation (\ref{eq:expand}) therefore compensates the noise
effects of the other local operations, leaving only the logical
function performed by the local operation corresponding to this
specific choice of input and output states. The quantum controlled-NOT
is thus capable of performing the logical functions of all three
local operations with a perfect fidelity of one, a task that cannot
be achieved by any positive sum of local conditional operations.

A quantitative criterion for the experimental observation of this
kind of quantum parallelism can be obtained from the fidelities of
the three classical logical operations associated with the local
operations $L_i$. These classical fidelities are defined as the
probability of obtaining the correct output, averaged over all four
possible inputs. In terms of the measurement probabilities
$P_{ij|kl}(a_{\mbox{\small out}} b_{\mbox{\small out}}
| a_{\mbox{\small in}} b_{\mbox{\small in}})$
of obtaining the logical output
$a_{\mbox{\small out}}$ and $b_{\mbox{\small out}}$ for a
given logical input $a_{\mbox{\small in}}$ and
$b_{\mbox{\small in}}$ in the logical operation realized
by choosing $k \otimes l$ eigenstates as input and measuring
$i \otimes j$ in the output, the fidelities of the three classical
logical operations are given by
\begin{eqnarray}
\label{eq:fidelities}
F_1 &=& \frac{1}{4} \left(P_{ZZ|ZZ}(00|00) + P_{ZZ|ZZ}(01|01)
+P_{ZZ|ZZ}(11|10) + P_{ZZ|ZZ}(10|11) \right),
\nonumber
\\[0.2cm]
F_2 &=& \frac{1}{4} \left(P_{XX|XX}(00|00) + P_{XX|XX}(11|01)
+P_{XX|XX}(10|10) + P_{XX|XX}(01|11) \right),
\nonumber
\\[0.2cm]
F_3 &=& \frac{1}{4} \left(P_{YY|XZ}(10|00) + P_{YY|XZ}(01|00)
+P_{YY|XZ}(00|01) + P_{YY|XZ}(11|01) \right.
\nonumber \\  && \hspace{0.5cm} \left.
+P_{YY|XZ}(00|10) + P_{YY|XZ}(11|10)
+P_{YY|XZ}(10|11) + P_{YY|XZ}(01|11) \right).
\end{eqnarray}
Here, the fidelity $F_1$ is simply the classical fidelity
of the controlled-NOT operation in the computational basis
($ZZ$-basis), as determined in previous experiments from the
truth table of the classical controlled-NOT operation.
Specifically, the measurement probabilities reported in
\cite{SKa03} correspond to an overall classical fidelity
of $F_1=73.5 \%$, and the classical fidelity reported in
\cite{Bri03} was $F_1=84 \%$. The determination of classical
fidelities is thus a straightforward and well established
experimental procedure for the characterization of
classical gate properties.
The fidelity $F_2$ for the classical controlled-NOT operation
observed in the $XX$-basis can be obtained by
testing the gate operation using the $XX$-basis instead of the
$ZZ$-basis to define both input states and output measurements.
Taken by itself, this fidelity is just another classical
characterization of the gate operation, without any indications
of quantum coherence or entanglement. However, only a quantum
gate can perform the controlled-NOT operation in both the $ZZ$-
and the $XX$-basis.
Finally, the third component of the ideal controlled-NOT operation
given in equation (\ref{eq:expand}) can be obtained by measuring
the truth table for an input in the
$YY$-basis and an output measurement of the $XZ$-basis. Here, each
input has two correct outputs, since the operation only defines the
correlation between output bits, not their specific individual values.
In the ideal case, each measurement probability contributing to
$F_3$ is therefore expected to be about $50 \%$.
A characterization of the quantum coherent properties of the gate
can thus be obtained by merely performing the classical evaluation
of individual gate operations for a selection of three different
input bases. As equation (\ref{eq:fidelities}) shows, this can
be achieved by recording the probabilities of 16 different
local measurement outcomes obtained with 12 different local
input settings.

The ideal quantum controlled-NOT gate is the only quantum process
that has perfect fidelities of $F_1=F_2=F_3=1$ for all three local
operations. On the other hand, the fidelities of the dephasing
operation $D$ are $F_1=F_2=F_3=1/2$. Each of the local operations
$L_i$ has one perfect fidelity of $F_i=1$ and two fidelities with
$F_j=1/2$ ($j \neq i$). Thus it can be conjectured that the average
fidelity for local operations is limited to a maximal value of
$2/3$ and that any expansion of the process into a sum of local
processes will require a negative dephasing component if this
limit is exceeded.
If the only source of errors is dephasing between the eigenstates
of $Z \otimes X$, it is possible to reconstruct the noisy
quantum controlled-NOT operation from the three fidelities $F_i$
by modifying the coefficients in the expansion given by equation (\ref{eq:expand}). The result reads
\begin{eqnarray}
\label{eq:estimate}
E_{\mbox{\small exp.}}(\hat{\rho}) &=& (2 F_1 - 1) L_1 (\hat{\rho})
+ (2 F_2 - 1) L_2 (\hat{\rho}) + (2 F_3 - 1) L_3 (\hat{\rho})
\nonumber \\
&& + 2 (2-F_1-F_2-F_3) D(\hat{\rho}).
\end{eqnarray}
In this expansion of a noisy quantum controlled-NOT operation,
quantum parallelism is expressed quantitatively in terms of
the contributions of each local operation $L_i$. Each of these
contributions is equal to $2 F_i-1$.
The number of parallel local operations effectively performed by
the gate can then be defined as the sum of the contributions
of the three operations $L_i$, given by $2 (F_1+F_2+F_3) - 3$.
Quantum parallelism is observed if this number is greater than one.
Thus the condition for quantum parallelism in
an experimental quantum controlled-NOT can be given by
\begin{equation}
\label{eq:criterion}
F_1+F_2+F_3 > 2.
\end{equation}
This means that the average fidelity of the three operations should be
greater than $2/3$ in order to verify quantum parallelism.
For lower average fidelities, equation (\ref{eq:estimate}) describes
a statistical mixture of local conditional operations that can be
performed without genuine quantum interactions, e.g. by using only local
operations and classical communication between the two qubits.

The most simple case of quantum process estimation is obtained
if all classical fidelities are equal. In this case, the quantum
gate can be described by a mixture of the ideal gate operation
$E_{\mbox{\tiny CNOT}}$ and the dephasing operation $D$,
\begin{eqnarray}
\label{eq:simple}
E_{\mbox{\small exp.}}(\hat{\rho}) &=&
p_{E} \; E_{\mbox{\tiny CNOT}}(\hat{\rho}) + (1-p_{E})
D (\hat{\rho})
\nonumber \\
&=& (2F-1) E_{\mbox{\tiny CNOT}}(\hat{\rho}) + 2 (1-F)
D (\hat{\rho}),
\end{eqnarray}
where $F_1=F_2=F_3=F$. The fidelity observed can then
be identified directly with the contribution $p_E$ of the ideal
operation $E_{\mbox{\tiny CNOT}}$ using $p_E=2F-1$. For instance,
when applied to an experimental fidelity of $75 \%$, this noise model
would suggest a quantum controlled-NOT contribution of $1/2$ and a
noise contribution of $1/2$. Nevertheless, this noisy operation
would still achieve quantum parallelism, since the fidelity is
greater than $2/3$.
Specifically, the condition for quantum parallelism in the
noisy operation given by equation (\ref{eq:simple}) is $p_E > 1/3$.
The role of the simplified noise model of equation (\ref{eq:simple})
for the evaluation of decoherence in quantum gates could thus be
similar to the role played by Werner states for the evaluation of mixed state entanglement \cite{Wer89}.

As mentioned above, the noise model defined by equation
(\ref{eq:estimate}) assumes that the only source of errors is
the loss of coherence between the eigenstates of $Z$ in system one
and $X$ in stystem two. This noise model has been chosen because
it represents the errors typically introduced by local simulations
of the quantum controlled-NOT operations, as given by the operations
$L_1$ to $L_3$. As will be discussed in more detail in the following,
it is thus most sensitive to the non-locality of the gate \cite{Har03}.
Specifically, a more general noise model will also include errors
that change the eigenvalues of $Z$ in system one and $X$ in stystem two.
However, such errors will reduce the fidelities $F_i$ more rapidly than
the dephasing errors represented by the local operations $L_i$.
Including such errors in the noise model for a given set of fidelities
$F_i$ would thus lead to a lower estimate for the total noise
and may cause an overestimation of the entanglement capability
of the gate.

It is in fact possible to proof that the criterion given by
inequality (\ref{eq:criterion}) provides an estimate of the
entanglement capability that is independent of the
noise model used. For this purpose, it is sufficient to
consider the amount of entanglement that can
be generated by an arbitrary noisy gate operation with fidelities
$F_i$. In order to relate these fidelities directly to the entanglement
capability, it should be noted that a classical
controlled-NOT operation will generate correlations between the target
and the control bit if the state of the control bit is random
and the input state of the target is known.
For example, a random mixture of the input states
$\mid Z=+1;Z=+1 \rangle$ and $\mid Z=-1;Z=+1 \rangle$ generates
output states with $\langle Z \otimes Z \rangle = 1$. However, the
density matrix of a random mixture of $Z$ eigenstates is $I/2$, the
same as that of a random mixture of $X$ eigenstates. Therefore, the successful
operation of the controlled-NOT in the Z-basis implies that a correlation
of $\langle Z \otimes Z \rangle = 1$ is also obtained from an input
state mixture of $\mid X=+1;Z=+1 \rangle$ and $\mid X=-1;Z=+1 \rangle$.
It is then possible to verify that the average magnitudes of the
correlations generated by applying the gate operation to
the four input states in the $XZ$-basis are related to the fidelities
$F_i$ by
\begin{eqnarray}
\label{eq:corr}
\overline{|\langle Z \otimes Z \rangle(\mbox{out})|} &\geq& 2 F_1-1
\nonumber \\[0.1cm]
\overline{|\langle X \otimes X \rangle(\mbox{out})|} &\geq& 2 F_2-1
\nonumber \\[0.1cm]
\overline{|\langle Y \otimes Y \rangle(\mbox{out})|} &=& 2 F_3-1.
\end{eqnarray}
These three correlations are sufficient to determine a lower bound
of the entanglement generated in the operation \cite{Hof03}. In
particular, the minimal concurrence $C$ corresponding to the correlations
in equations (\ref{eq:corr}) is given by
\begin{eqnarray}
C &\geq& \frac{1}{2}\left(|\langle X \otimes X \rangle|+
|\langle Y \otimes Y \rangle|+|\langle Z \otimes Z \rangle|-1 \right)
\nonumber \\
&\geq& F_1+F_2+F_3-2.
\end{eqnarray}
This result confirms the criterion for quantum parallelism given by
inequality (\ref{eq:criterion}). In fact, equation (\ref{eq:corr})
indicates that it is even possible to identify the precise contribution
of each local operation to the inseparable correlations of the entangled
output state. The intuitive notion that quantum
parallelism corresponds to a simultaneous performance of distinct
local operations expressed by the decomposition in equation
(\ref{eq:estimate}) is thus confirmed by the possibility of generating
entanglement when the fidelity limits of local operations
are exceeded.

In previous tests of experimental quantum gates, the verification of
entanglement generation has been performed separately from the
determination of the classical fidelity $F_1$ in the computational basis \cite{SKa03,Bri03}. The results given above show that a more consistent
evaluation of classical fidelities and entanglement capability can be
achieved by measuring the complete set of three classical
fidelities $F_i$. By fully characterizing the essential operations
of the quantum controlled-NOT, the fidelities $F_i$ also provide a
measure of how closely any experimental realization approximates
the ideal quantum gate.
Such a measure has only been given in \cite{Bri04}, where it is noted
that the measurement probabilities of 65 local settings of input
and output states were necessary to evaluate the process fidelity.
In contrast, the proposed evaluation
of quantum parallelism requires only the 16 measurement probabilities
needed to determine the fidelities $F_i$
according to equation (\ref{eq:fidelities}).
In the light of the discussions given in \cite{SKa03,Bri03,Bri04},
it seems that the fidelities $F_i$ can provide a
surprisingly compact characterization of the essential quantum gate properties.


It should also be noted that the present approach not only provides
a measure of gate performance, but also identifies and
characterizes three different functions of the gate which,
when combined, make up the complete operator of the gate, $E_{\mbox{\tiny{CNOT}}}$. As the initial derivation of the local processes
shows, each of the three fidelities can be identified with a
specific coherence in the process matrix given in equation (\ref{eq:qcNOT}). The particular choice of the three fidelities
is thus determined by the expansion of the gate operation into
four locally defined components. It may be possible to characterize
the gate using even less measurements, but such procedures would
not evaluate the complete quantum coherence, or quantum parallelism,
of the operation. Likewise, it is possible to obtain a more
detailed insight into the noise signature of the operation by
measuring more classical fidelities. However, the three fidelities
given here already determine the essential quantum coherences
defining the quantum parallelism of the operation. In general,
the procedure introduced above thus identifies the essential quantum
parallelism of multi qubit gates with a minimal set of representative
classical fidelities.
Specifically, an average fidelity below
$2/3$ indicates that the performance of the gate can be
reproduced by local operations and classical communications according
to the decomposition given in equation (\ref{eq:estimate}),
while an average fidelity above $2/3$ proofs that no such local
decomposition exists.

In conclusion, it has been shown that the quantum coherent
operation of an ideal quantum controlled-NOT can be expressed
in terms of the parallel performance of three distinct local
operations. Each of these local operations $L_i$ corresponds
to a characteristic logical
function that can be evaluated experimentally by measuring its
classical fidelity $F_i$. If the average fidelity of the three
operations exceeds $2/3$, the experimental gate effectively
performs more than one local operation in parallel and
entanglement generation is possible. An estimate of the noisy
quantum operation can also be obtained by adjusting the
statistical weight of the operations $L_i$ according to the
observed fidelities $F_i$.
The measurement of the three classical fidelities $F_i$ thus
provides an efficient test of the performance of experimental
quantum controlled-NOT gates.

Part of this work has been supported by the JST-CREST project on
quantum information processing.


\begin{thebibliography}{xyz00}
\bibitem{schmidt}
Note that this decomposition of a two qubit gate is the formal equivalent
of the Schmidt decomposition of entangled quantum states.
\end{thebibliography}
\end{document}